\documentclass[aps,pra,showpacs,superscriptaddress]{revtex4}

\usepackage{amsmath}
\usepackage{amsfonts}
\usepackage{amssymb}

\newtheorem{theorem}{Theorem}
\newtheorem{lemma}{Lemma}

\def\bra#1{\langle #1 |}
\def\ket#1{| #1\rangle}

\begin{document}

\title{
Security of Quantum Bit-String Generation}

\author{Jonathan Barrett}
\email{jbarrett@ulb.ac.be}
\affiliation{Physique Th\'{e}orique, {C.P.} 225, Universit\'{e} Libre
de Bruxelles, Boulevard du Triomphe, 1050 Bruxelles, Belgium} 
\affiliation{Centre for Quantum Information and Communication, {C.P.}
165/59, Universit\'{e} Libre de Bruxelles, Avenue F. D. Roosevelt 50,
1050 Bruxelles, Belgium}  
\author {Serge Massar}
\email{smassar@ulb.ac.be}
\affiliation{Physique Th\'{e}orique, {C.P.} 225, Universit\'{e} Libre
de Bruxelles, Boulevard du Triomphe, 1050 Bruxelles, Belgium} 
\affiliation{Centre for Quantum Information and Communication, {C.P.}
165/59, Universit\'{e} Libre de Bruxelles, Avenue F. D. Roosevelt 50,
1050 Bruxelles, Belgium}  

\begin{abstract}
We consider the cryptographic task of bit-string generation. This is a generalisation of coin tossing in which two mistrustful parties wish to generate a string of random bits such that an honest party can be sure that the other cannot have biased the string too much. We consider a quantum protocol for this task, originally introduced in Phys. Rev. A {\bf 69}, 022322 (2004), that is feasible with present day technology. We introduce security conditions based on the average bias of the bits and the Shannon entropy of the string. For each, we prove rigorous security bounds for this protocol in both noiseless and noisy conditions under the most general attacks allowed by quantum mechanics. Roughly speaking, in the absence of noise, a cheater can only bias significantly a vanishing fraction of the bits, whereas in the presence of noise, a cheater can bias a constant fraction, with this fraction depending quantitatively on the level of noise. We also discuss classical protocols for the same task, deriving upper bounds on how well a classical protocol can perform. This enables the determination of how much noise the quantum protocol can tolerate while still outperforming classical protocols. We raise several conjectures concerning both quantum and classical possibilities for large $n$ cryptography. An experiment corresponding to the scheme analysed in this paper has been performed and is reported elsewhere.

\pacs{03.67.-a 03.67.Dd}

\end{abstract}

\maketitle

\section{Introduction}

Coin tossing is a cryptographic primitive introduced by
Blum \cite{blum1982}, in which two parties who do not trust one
another want to agree on a random 
bit. An honest party must be sure that the other party cannot have biased the bit if they cheated. Such protocols can be divided into two classes, according to
whether the parties know or do not know beforehand which value of the
coin the other party desires. These are known respectively as weak and strong coin
tossing. Classically, these tasks can be achieved if assumptions are made that
limit the computational power of a dishonest party \cite{blum1982},
or if relativistic signalling constraints are used
\cite{kentrel}. They can also be implemented using a trusted source of noise or a trusted third party. Without such assumptions, however, one of the
parties can always fix the value of the coin with certainty if he or
she cheats. 

If the parties can use quantum communication,
then  non-trivial protocols exist with security guaranteed by the
laws of quantum mechanics. In the case of strong coin tossing, this was first
shown by Aharonov \emph{et al.} \cite{aharonovta-shmavaziraniyao2000}. The best strong coin tossing
protocol to date is due to Ambainis \cite{ambainisnewprotocol2002}, and independently to Spekkens and Rudolph \cite{spekkensrudolph}; the bias achieved is $1/4$. Weak coin tossing was first considered by Goldenberg \emph{et al.} in the context of quantum gambling \cite{GVW}, and was subsequently generalised by Spekkens and Rudolph \cite{spekkensrudolphcheatsensitive2002}. On the other hand, it was first shown by Lo and Chau that coin tossing (weak or strong) with perfect security is not possible \cite{lochaucointoss}.
Subsequently, a lower bound on the achievable bias for strong coin
tossing was proven by Kitaev \cite{Kitaev}.
No further bounds on weak
coin tossing are known, although it is known that the smaller the bias
the more rounds of communication are required \cite{ambainisnewprotocol2002}. 

Coin tossing is often introduced via an example of two parties who have divorced and want to decide who gets the car. Its real importance, however, lies in the fact that it is a useful primitive for the construction of more general cryptographic protocols. Blum, in his original work \cite{blum1982} notes that in the classical context, coin tossing can be used to implement mental poker and certified mail. More recently, 
Kent has suggested that by building on secure coin tosses, it may be possible to construct quantum ``classically certified bit commitment'' (hence oblivious transfer and general secure multi-party computation), with security based on the hardness of an NP-complete problem \cite{kentuses}. A scheme with this type of security, it is widely conjectured, would be secure against any polynomial-time quantum attack. 

When many coins are being tossed, rather than a single one, we call this ``bit-string generation''. Most applications will clearly involve bit-string generation, rather than a single coin toss.
It may seem as if the question of whether bit-string generation can be made secure should reduce trivially to the question of whether single-shot coin tossing can be secure. In general, however, the security of large $n$ cryptography does not reduce simply to the security of the single-shot case. For example, it may be possible to attain a certain level of security for the entire string, even though individual bits of the string are not secure. This was
pointed out by Kent, in the context of quantum bit-string commitment \cite{kentbitstringcommitment}. Kent has also discussed bit-string generation \cite{kent}. He introduces a quantum protocol for bit-string generation and argues that his protocol gives good security in the case of no noise and large $n$, although does not provide a detailed analysis. 

A different protocol for bit-string generation was introduced in
Ref.~\cite{BM}, which has the advantage of being feasible with present
day technology. A security analysis was given that applies in the
realistic case that the quantum channel separating the two parties is
noisy. The analysis, however, had two drawbacks: first, the security
condition adopted, the so-called average bias condition, is not very
restrictive, and second, only a limited class of attacks (individual
attacks) were considered. In this work, we build on the results of Ref.~\cite{BM}. We introduce
a new, stronger security condition based on the Shannon entropy of the
string. Using both the average bias condition and the Shannon entropy
condition, we consider the security of the protocol in the absence and in the
presence of noise, under the most general attacks allowed by quantum
mechanics. We give rigorous proofs 
that in the absence of noise, the protocol has good
security, where roughly speaking this means that a cheater can only
fix the values of a vanishing fraction of the coins. In the presence
of noise, the protocol is partially secure, with the level of security
depending quantitatively on the level of noise.  

Noise can of course be counteracted using quantum error
correction codes or entanglement distillation, and in principle be
reduced to an arbitrarily low level. Nonetheless there are at least
two good reasons for including noise in the analysis. One is that
cheating strategies are in general indistinguishable from noise in the
communication channel (as far as the honest party is concerned), and
it is therefore very natural to carry out the theoretical analysis in
this case, rather than in the noiseless case.  
Another reason is that we wish our results to apply to the present day
experimental situation, which does not allow for the reliable
implementation of quantum error correction codes or entanglement
distillation. 

The ultimate measure of success for a quantum cryptographic protocol
must be whether the protocol gives security that is adequate for use
in a real practical situation. The level of security required will
determine the degree of noise that can be tolerated, and will
obviously depend on the circumstances. It is possible that
technological improvements will be required before this level can be
reached. In the meantime, a useful figure of merit is whether the
quantum protocol is achieving a level of security that cannot be
obtained classically. Thus it is important to contrast classical protocols
for the same task. With this motivation, we shall also discuss purely
classical protocols for bit-string generation, under various security
conditions. We derive some bounds on how well classical protocols can
perform. We also give an interesting example of how the problem of the
best classical protocol is not always trivial. 

Using these classical bounds, it is possible to show that our quantum protocol, if implemented
with present day technology, can achieve a level of security that is
impossible classically. We report elsewhere \cite{Expt} an experimental
realisation of quantum bit-string generation, based on the protocol
and security analysis presented here. We note
that another experiment realising quantum coin tossing has recently
been reported \cite{Expt2}. This experiment is an impressive
achievement from the point of view of physics (for example, it is one
of the first to realise individual control over quantum qutrits).
In contrast with the experiment
of Ref.~\cite{Expt}, however, the security analysis is incomplete; in fact, it
is not clear that anything classically impossible has been achieved. 

We shall begin in Sec.~\ref{securitysection} by defining the task of
bit-string generation, along with some precise security conditions. In
Sec.~\ref{classicalsection}, we investigate classical protocols for
bit-string generation, proving some bounds on the level of security
that can be achieved. In Sec.~\ref{protocolsection}, we introduce our quantum
protocol and discuss briefly the most general attacks available to a
dishonest party, before presenting our main results in
Sec.~\ref{resultssection}. The proofs of these results are given in
Sec.~\ref{proofssection}. Finally, Sec.~\ref{discussionsection} contains some further
discussion.

\section{Security conditions}\label{securitysection}

In this work, we do not assume any restrictions on the computational power of an adversary. Neither are there any trusted sources of noise or third parties. We assume a non-relativistic scenario (this means that there is no way of ensuring a simultaneous exchange of messages, thus a protocol can only involve a sequential exchange). Two parties, Alice and Bob, are assumed to occupy separated laboratories. A dishonest party is assumed to have control over everything outside the honest party's laboratory. In the quantum case, we are interested in unconditional security, meaning that a dishonest party is limited only by the laws of quantum mechanics. In the classical case, we are interested in information theoretic security.

A coin tossing or bit-string generation protocol consists of a sequence of rounds of communication between Alice and Bob. In the quantum case, the communication may of course be quantum, and local operations such as adding ancillas or performing measurements may be carried out at any stage. For precise security conditions for single-shot coin tossing, we refer the reader to Ref.~\cite{BM}. Here we consider only bit-string generation. Thus consider a protocol in which two mistrustful parties, Alice and
Bob, want to toss $n$ coins. When the protocol terminates, Alice either outputs an $n$-bit string ${\bf x}$, or she is deemed to have aborted the protocol, in which case we write ${\bf x} = \perp$. Similarly, Bob either outputs an $n$-bit string ${\bf y}$, or aborts, in which case ${\bf y}= \perp$. Roughly speaking, a good protocol should ensure that a cheating Alice cannot bias Bob's output too much and vice versa. We emphasise that throughout this work, we are interested in two-party protocols, meaning that Alice and Bob are mistrustful and it is they who may be dishonest. We are not concerned with the possibility of dishonest third parties or eavesdropping, and there is no requirement of secrecy.

In the ideal case, we should demand that when both
parties are honest, the protocol never aborts, ${\bf x}={\bf y}$, and the coins are all fair.  We express this as
\begin{equation}
\forall {\bf c}\in\{0,1\}^n\quad 
 \mathrm{P}^{H_A H_B}({\bf x}={\bf y}={\bf c}) = 2^{-n}, 
\end{equation}
where $H_A$, $H_B$ denote the honest strategies of Alice and Bob. In any real implementation there will be some finite noise level, and so this condition will not hold exactly since, due to the noise, there will be a small chance of aborting even when both parties are honest. We replace this condition, therefore, with a slightly weaker one. We say that a protocol is \emph{correct} if
\begin{equation}\label{noisycorrectness}
\forall {\bf c}\in\{0,1\}^n\quad\quad \frac{1 - \delta_n}{2^{n}} \leq
 \mathrm{P}^{H_A H_B}({\bf x}={\bf y}={\bf c}) \leq \frac{1 + \delta_n}{2^{n}}, 
\end{equation}
where we demand that $\delta_n$ tends to zero as $n$ (or indeed some other parameter of the protocol) increases. 

It is possible to think of many different measures of the security of
a bit-string generation protocol. Here, we focus on three main types of
security.  

{\bf Average Bias.}
We denote by $S_A$ and $S_B$ arbitrary strategies of Alice and
Bob. Then we define
the \emph{average bias} for each by
\begin{eqnarray}
\frac12 + \epsilon_A &=& \max_{S_A, {\bf c}\in\{0,1\}^n}
\ \frac{1}{n} \sum_{i=1}^n  \mathrm{P}^{S_A H_B}(y_{i}=c_i),\label{defEA}\\
\frac12 + \epsilon_B &=& \max_{S_B, {\bf c}\in\{0,1\}^n}
\ \frac{1}{n} \sum_{i=1}^n  \mathrm{P}^{H_A S_B}(x_{i}=c_i),\label{defEB}
\end{eqnarray}
where $x_i$ is the $i$th bit of ${\bf x}$ and $y_i$ the $i$th bit of ${\bf y}$.

{\bf Shannon Entropy}
The average bias is a simple measure of security but it is not very
satisfactory. Consider, for example, the case in which Alice can cheat so that Bob's output is either the string composed of all zeros or the string composed of all
ones, with equal probabilities: $\mathrm{P}({\bf
y}=0^n)=\mathrm{P}({\bf y}=1^n)=1/2$. The average bias $\epsilon_A$ is zero,
although the security is clearly very bad. 

For this reason we introduce another security condition based on the 
\emph{Shannon entropy} of the string. For the purposes of this condition it is convenient to assume that an honest party never aborts. If, for example, Bob gains evidence that Alice is cheating, and the protocol stipulates that he should abort, we assume instead that he chooses an $n$-bit string randomly and independently from the rest of the protocol, and outputs that. Similarly Alice. The protocol should ensure that a cheating party cannot reduce the entropy of the other party's output too much. \footnote{Another approach would be to allow the abort outcome, to stipulate that we ignore any strategy that aborts with probability exponentially close to $1$, and then to focus on the entropy of the string conditioned on the protocol not being aborted. This would be similar to the manner in which security of quantum key distribution is usually approached. Our results could easily be adapted to this approach, but would take a more complicated form.}

We define \footnote{The symbols $H_A$ and $H_B$ defined here are identical to those used to denote Alice's and Bob's honest strategies. However, it will always be clear which is meant.}
\begin{eqnarray}
H_A &=& \min_{S_{A}} H\left(\mathrm{P}^{S_{A},H_B}({\bf y})\right), \label{defHA}\\
H_B &=& \min_{S_{B}} H\left(\mathrm{P}^{H_A,S_{B}}({\bf x})\right)\label{defHB},
\end{eqnarray}
where $H$ is the usual Shannon entropy of a probability distribution, i.e.,
\begin{equation}
H\left(\mathrm{P}({\bf x})\right) = -\sum_{{\bf x}} \mathrm{P}({\bf x})\log\mathrm{P}({\bf x}).
\end{equation}
(Here and throughout this work, $\log$ denotes a logarithm of base 2.)

Now we say that a bit-string generation protocol is \emph{arbitrarily secure} if
$n - H_A \rightarrow 0$ and $n - H_B \rightarrow 0$ as
$n\rightarrow \infty$. Similarly, it is \emph{relatively secure}
if $(n - H_A)/n \rightarrow 0$ and $(n-H_B)/n \rightarrow 0$ as
$n\rightarrow \infty$. Roughly speaking, this means that a cheater may be able to fix the values of some of the coins, but that the fraction of coins thus affected must become small as $n$ increases. It is \emph{partially secure} if
$H_A,H_B > 0$. Our main results will be that our quantum protocol is relatively secure in the absence of noise and partially secure in the presence of noise, with security depending quantitatively on the amount of noise.

{\bf Min-entropy.}
Finally, we introduce a security condition based on the maximal
probability of occurrence of a string (referred to as
the min-entropy condition) 
\begin{eqnarray*}
H^\infty_A &=& 
- \log_2\ \max_{S_A, {\bf c}\in\{0,1\}^n}\ \mathrm{P}^{S_A H_B} ({\bf
y} = {\bf c}),\\ 
H^\infty_B &=& 
- \log_2\ \max_{S_B, {\bf c}\in\{0,1\}^n}\ \mathrm{P}^{H_A S_B} ({\bf
x} = {\bf c}). 
\end{eqnarray*}
For the purposes of this condition, we again allow honest parties to abort.

We will not discuss the min-entropy condition in much detail in this work, because the quantum protocol we study does not give good security with respect to this condition. In principle, one could define arbitrary, relative, and partial security in terms of the min-entropy condition, in a manner precisely analogous to their definition in terms of the Shannon entropy. It turns out, however, that even in the absence of noise the protocol is not then relatively secure. 
This should be contrasted with the other security conditions we
defined above for which we show that good security can be achieved in
the absence of noise. We note that the bit-string generation protocol due to Kent \cite{kent} does not achieve relative security with respect to the min-entropy condition either, and conjecture that no quantum protocol can do so. We have introduced the min-entropy here because we are able to
prove a bound on the achievable min-entropy by any classical protocol.

From the above security conditions, we can see that the relationship
between coin tossing and bit-string generation is not trivial. A good
protocol for coin tossing, for example, does not necessarily imply a
good protocol for bit-string generation. A perfectly secure coin
tossing protocol that is simply repeated many times will result in a
bit-string generation protocol that satisfies our average bias
condition above, with $\epsilon_A=\epsilon_B=0$. But it will not
necessarily be arbitrarily or relatively secure unless the coin
tossing protocol is composable. If it is a quantum protocol, one would
have to consider the possibility that it is not composable because a
cheater can entangle separate runs. A bit-string generation protocol
that is arbitrarily secure does imply a coin tossing protocol - simply
take the first bit of the string. However, a bit-string generation
protocol that is relatively secure need not. From this, and Kitaev's
lower bound for quantum coin tossing, we can conclude that quantum
bit-string generation with arbitrary security is not possible. In this
work, we therefore consider mainly relative security (for the
noiseless case) and partial security (for the noisy case). 

\section{Classical bit-string generation}\label{classicalsection}

As stated above, classical coin tossing (both weak and strong) is impossible with information theoretic security. It turns out that at least one party can fix the outcome with certainty (for a proof, see, e.g., Theorem~2 of Ref.~\cite{Kitaev}). There is, however, a trivial protocol for
bit-string generation that achieves partial security: assuming even
$n$, Alice tosses half of the coins herself, and sends the results to
Bob, who then tosses the other half and sends the results to Alice. In
this section we show that this trivial protocol is optimal among
classical protocols, both with respect to average bias and with
respect to min-entropy. Then we discuss classical protocols and the Shannon entropy condition. 
\begin{theorem}\label{theorem1}
For any conceivable classical protocol,
\[
\epsilon_A + \epsilon_B \geq 1/2.
\]
The trivial protocol saturates this bound.
\end{theorem}
{\bf Proof}
The theorem follows directly from the impossibility of classical single-shot
coin tossing, since for each $i$, at least one of $\mathrm{P}^{S_AH_B}(y_i=c_i)$ and $\mathrm{P}^{H_AS_B}(x_i=c_i)$ can be made equal to $1$ or $0$ by a cheater. \quad $\Box$

We can also prove a bound on the classically achievable min-entropy:
\begin{theorem}\label{theorem2}
For any conceivable classical protocol,
\[
\min[H^\infty_A,H^\infty_B] \leq n/2.
\]
It is clear that the trivial protocol saturates this bound.  
\end{theorem}
{\bf Proof}
In this case, a proof follows from what is essentially Theorem~1 of
Ref.~\cite{Kitaev}, but generalised to the case of bit-string
generation. For completeness, we include the generalised version of
the proof here. The presentation is very similar to that of Ref.~\cite{Kitaev}.

Assume that a classical protocol involves $k$ rounds of communication 
\footnote{
In general, one should also consider protocols for which the number of
rounds is not fixed but depends on random choices made during the
execution. Such protocols may even have an unbounded number of rounds,
as long as the average is finite. Our proof would have to be
generalised to cover this case.}.  
Let $U$ denote the state of the protocol at any given moment. Thus $U$
contains a specification of all communications sent by Alice and Bob
up to that particular point. We define $w(U)$ to be the probability of
state $U$ occuring during an honest execution of the protocol. We
define two other functions of $U$. Let 
\[
Z^A_{{\bf c}}(U) \equiv \max_{S_B}\,\mathrm{P}({\bf x}={\bf
c}|\mathrm{\ state\ is\ }U), 
\]
where the maximum is over all strategies that Bob can employ from the
point $U$ onwards. Similarly, 
\[
Z^B_{{\bf c}}(U) \equiv \max_{S_A}\,\mathrm{P}({\bf y}={\bf c}
|\mathrm{\ state\ is\ }U). 
\]
Finally, let
\[
F_j = \sum_{U\in {\cal U}_j} w(U)\,Z^A_{{\bf c}}(U)\,Z^B_{{\bf c}}(U),
\]
where ${\cal U}_j$ is the set of those $U$ that specify a state of the
protocol after $j$ rounds of communication. It is not too difficult to show that
$F_j\geq F_{j+1}$. This allows us to conclude that 
\begin{equation}
\max_{S_A}\left[\mathrm{P}^{S_A,H_B}({\bf y}={\bf c})\right] \times
\max_{S_B}\left[\,\mathrm{P}^{H_A,S_B}({\bf x}={\bf c})\right] =
Z^B_{{\bf c}}(U_0)\,Z^A_{{\bf c}}(U_0)=F_0\geq
F_{k}=\mathrm{P}^{H_A,H_B}({\bf x}={\bf y}={\bf c})=1/(2^n), 
\end{equation}
where $U_0$ is the state of the protocol before the first
communication. This gives us Theorem~\ref{theorem2}. \quad $\Box$

These two results show that there are rather strong limitations on
what classical protocols can achieve. Theorem~\ref{theorem1} is particularly useful as it enables us to determine precisely for what levels of noise our quantum protocol beats this bound, and thus for what levels of noise something classically impossible is being achieved.

Finally, it may appear from the above that there are no interesting
classical possibilities beyond the trivial protocol. We conclude this
section with a diverting counterexample: if we adopt our security criterion based on Shannon entropy, then
there are classical protocols that outperform the trivial protocol, at
least for finite $n$. It is clear that the trivial protocol gives
$H_A,H_B=n/2$. Now consider the following. Alice sends a
communication to Bob that specifies a particular $n$-bit string. This
string will not be the outcome of the protocol. Bob then sends a
communication to Alice that rules out another bit-string. This
continues until all bit-strings have been ruled out except one, which
is the outcome of the protocol. It can be shown that in the
case of four bits (sixteen strings), we have $H_A\approx 2.39 >
2$ and $H_B\approx 2.78 > 2$. Thus we have improved on the
trivial protocol both from the point of view of cheating Alice and
cheating Bob. The bounds above, however, must apply, so this protocol
does not improve on the trivial protocol with respect to average bias
or min-entropy.  
Interestingly, numerical investigations indicate that for this
protocol, $(n-H_A)/n$ and $(n-H_B)/n\rightarrow 1/2$ as $n \rightarrow
\infty$. If correct, this means that the advantage disappears in the
large $n$ limit. 

A bound on the achievable $H_A$ and $H_B$ by any classical protocol would be useful. Combined with our results on Shannon entropy for our quantum protocol below, it would enable us to determine for what noise levels the quantum protocol is outperforming all classical protocols with respect to this condition. Our results for the protocol described in the last paragraph lead us to conjecture that, with respect to Shannon entropy, no classical protocol outperforms the trivial protocol in the large $n$ limit (more precisely, we conjecture that for any classical protocol, $H_A + H_B \leq n + o(n)$, where $o(n)$ denotes a term such that $o(n)/n \rightarrow 0$ as $n\rightarrow \infty$). As things stand, however, one can at least use our results for the average bias condition to determine a rigorous quantum-classical separation. This was our main reason for including these results.    

\section{A quantum bit-string generation protocol}\label{protocolsection}

\subsection{The protocol}

We now describe the quantum protocol for which we will prove security bounds.
\\ \\
{\bf Protocol 1:} Denote by $n$ the length of the bit-string to be generated. 
Let $|\psi_0\rangle$ and $|\psi_1\rangle$ be two non-orthogonal quantum
states with $|\langle \psi_0|\psi_1\rangle|^2 = \cos^2 \theta$. In general, $\theta$ may be fixed or may be a function of $n$. However, both $n$ and $\theta$ are fixed before the commencement of the protocol. Fix also $0<f^*\leq 1$.  

\begin{enumerate}
\item For $i=1$ to $n$
\begin{enumerate}
\item Alice chooses a random bit $a_i\in\{0,1\}$. 
She prepares the quantum state
  $|\psi_{a_i}\rangle$. She sends $|\psi_{a_i}\rangle$ to Bob.
\item Bob chooses a random bit $b_i\in\{0,1\}$. Bob reveals $b_i$ to Alice.
\item Alice reveals $a_i$ to Bob.
\item Bob measures the state sent to him by Alice using a two outcome von
  Neumann measurement which either projects onto
  $|\psi_{a_i}\rangle\langle\psi_{a_i}|$, or onto the orthogonal subspace
$I - |\psi_{a_i}\rangle\langle\psi_{a_i}|$. 
If the outcome of the measurement corresponds to
$|\psi_{a_i}\rangle\langle\psi_{a_i}|$, then $f_i=1$; 
  if the outcome of the measurement corresponds to the orthogonal subspace then $f_i=0$.
\end{enumerate}
\item Next $i$
\item Alice outputs ${\bf x}$, where $x_i=a_i\oplus b_i$.
\item If $\sum_{i=1}^n f_i \geq n f^*$, then Bob outputs ${\bf y}$, where $y_i = a_i \oplus b_i$. 
\item If $\sum_{i=1}^n f_i < nf^*$ then Bob aborts and ${\bf y}=\perp$.
\end{enumerate}

We specify in addition that an honest party should always abort if it is clear that the other party has failed to follow their part of the protocol, e.g., if an expected classical bit never arrives. 

In the absence of noise, the constant $f^*$ in Protocol~1 can be taken
to be equal to 
$1$. In the presence of noise, however, even if Alice and Bob are both honest,
there is a finite probability that Bob's measurement will fail (and
therefore that
$f_i=0$). This means that if $f^*=1$, then the
probability that the protocol does not abort is exponentially
small, and the protocol is not correct. As we argue below, however, by
choosing $f^*$ sufficiently small, we have that
$\delta_n$ in Eq.~(\ref{noisycorrectness}) tends to zero exponentially
fast as $n$ tends to infinity. Thus correctness is satisfied.  

\subsection{Cheating and noise}\label{cheatingsection}
One of the aims of this work is to consider how to carry out quantum
bit-string generation 
in the presence of noise. Thus we shall consider the security of the above protocol, both in the absence and the presence of noise, under the most general attacks allowed by quantum mechanics. We suppose that the noise
manifests itself as an imperfect quantum communication channel between Alice
and Bob. Such a channel can always be modelled as
$|\phi\rangle\langle\phi|\rightarrow {\cal
S}(|\phi\rangle\langle\phi|)$, where $|\phi\rangle\langle\phi|$ is an
input to the channel and ${\cal S}$ is a completely positive
trace-preserving map. One could also consider the effect of
imperfections in Alice's 
and Bob's laboratories, such as finite detection efficiencies or
imperfect state preparation procedures. We will not take these into
account here, and refer to Ref.~\cite{Expt} for this more general
case. We shall always assume that classical channels are noiseless. 

In the presence of noise, we adopt the most pessimistic assumption,
which is that a dishonest party can in principle replace the noisy
communication channel by a perfect channel. Then, as long as the
cheating is not excessive, it will not lead the other party to abort,
since any errors induced will be indistinguishable 
from the expected noise. The situation is similar to that which arises
in quantum key distribution, where the presence of noise is
indistinguishable from the presence of an eavesdropper. 

The most general attack for a cheating Alice is to replace the noisy channel with a noiseless channel and then to prepare $n+1$ systems in some joint, possibly entangled, state. Each round, she sends one system to Bob. After Bob sends $b_i$, she performs a positive operator-valued (POV) measurement on the systems left in her possession, which in general may depend on the value of $b_i$, and indeed on events in the previous rounds. The outcome of this measurement, along with previous events, will determine the value of $a_i$. At any point, she may simply decide to stop following the protocol, thus causing Bob to abort, although we shall see below that this latter strategy cannot help. 

The most general attack for a cheating Bob is to replace the noisy channel with a noiseless channel, as desired, and then to measure each quantum state sent by Alice as soon as it arrives (i.e., before sending the bit $b_i$). The bit $b_i$ may then depend on the outcome of this measurement. Clearly Bob can correlate his strategy over different rounds if he wishes, although, as we argue below, this will not be of use. Bob can also decide to stop following the protocol, thus causing Alice to abort, but again, this cannot help.

\section{Results}\label{resultssection}

We state our main results in the form of a set of bounds that concern the security of Protocol 1 in both noisy and noiseless conditions. In this section, we give the results in a simple asymptotic form. Exact results can be found in the proofs below.

\begin{theorem}\label{noiselesstheorem}
Noiseless case. Set $f^*=1$ and $\sin^2\theta=(\ln n)^{1/6} n^{-1/6}$. Then the
protocol is correct and we have 
\begin{equation}
\max\{\epsilon_A , \epsilon_B \} \leq O\left(\frac{\ln
n}{n}\right)^{1/12}.\label{noiselesseaeb}
\end{equation}
Alternatively, if we set $f^*=1$ and $\sin^2\theta=(\ln n)^{1/8}n^{-1/8}$, we get
\begin{equation}
\min\{ H_{A},H_{B} \} \geq n - O((\ln n)^{1/8}n^{7/8}).\label{noiselesshahb} 
\end{equation}
Thus the protocol is relatively secure and is better than any
classical protocol. 
\end{theorem}
Taking $\sin\theta$ to be a decreasing function of $n$ is the key to
obtaining Eqs.~(\ref{noiselesseaeb}) and (\ref{noiselesshahb}). When $\sin\theta$ decreases it is harder
and harder for Bob to cheat, since it is harder and harder for him to
guess the state sent to him by Alice. On the hand it is easier and
easier for Alice to cheat, since the states she must send Bob are more
and more similar. There is an optimal rate of decrease of $\sin\theta$ which balances these two effects. 

\begin{theorem}\label{noisytheorem}
Noisy case. Fix $f^*$ such that it is smaller than the fidelity of the
quantum channel \footnote{Strictly speaking, we mean the fidelity averaged over the two states sent by honest Alice: $1/2 \bra{\psi_{a_0}}{\cal E}(\ket{\psi_{a_0}})\ket{\psi_{a_0}} + 1/2 \bra{\psi_{a_1}}{\cal E}(\ket{\psi_{a_1}})\ket{\psi_{a_1}}$, where ${\cal E}$ denotes the map corresponding to the channel. For brevity, we refer to this as 'the' fidelity throughout.} . Fix $\theta$ independently of $n$. Then the protocol
is correct and we get 
\begin{eqnarray}
\epsilon_A &\leq&
\frac{\sqrt{(1-f^*)}}{\sqrt{2}\sin^2\theta} + \frac{1-f^*}{\sin^2\theta}+ O\left(\sqrt{\frac{\ln
n}{n}}\right),\label{noisyea}\\ 
\epsilon_B &\leq& \frac{\sin \theta}{2},\label{noisyeb}\\
H_{A} &\geq& n \left( - \log\left[\frac12+\frac{\sqrt{1-f^*}}{\sqrt{2}\sin^2\theta}+\frac{1-f^*}{\sin^2\theta}\right]\right) - O(\sqrt{n\ln n}),\label{noisyha}\\ 
H_{B} &\geq& n h\left(\frac12(1+\sin\theta)\right),\label{noisyhb}
\end{eqnarray}
where in the last line, $h$ is the binary entropy function, $h(p)=-p\log p - (1-p)\log (1-p)$.
The protocol is partially secure. For sufficiently low noise, we can set
$f^*$ such that it is better than any classical protocol. 
\end{theorem}

\section{Proofs}\label{proofssection}
\subsection{Correctness Condition}

One easily checks that if both parties are honest, and if the protocol does not abort, then the coins are fair. If both parties are honest, and if the fidelity of the communication channel is $f^0> f^*$, then a standard result in probability theory implies that the probability that the protocol aborts decreases exponentially with $n$:
\begin{equation}
\delta_n \leq \exp[-n\frac{(f^0-f^*)^2}{2}].
\end{equation}
Hence the protocol is correct.

\subsection{Dishonest Bob}

In order to cheat, Bob measures on each round the state sent by Alice. He does this before announcing $b_i$. His aim is to guess correctly the value of $a_i$, and then choose the value of $b_i$ so as to obtain the outcome for that particular coin that he wants. Recalling that a cheating Bob can replace the noisy channel with a perfect channel, his task is therefore to perform a measurement that distinguishes as well as possible the two non-orthogonal states $|\psi_0\rangle$ and $|\psi_1\rangle$.  A standard result in state estimation \cite{Helstrom} states that the probability that Bob guesses correctly is bounded by
\begin{equation}
\mathrm{P(correct\ guess)} \leq \frac12 +\frac{\sin \theta}{2}.
\label{EB}
\end{equation}

When analysing the security with respect to a dishonest Alice, a big
complication is that in principle, Alice can make her strategy at round
$i$ depend on what happened during the previous rounds. On the other hand, in 
the case of Bob, correlating his strategy at one
round with the strategy at previous rounds cannot help. This is because at each
round the state sent by Alice is chosen at random, independently from the rest of the protocol. At each round, if Bob performs any measurement other than the optimal distinguishing measurement of Eq.~(\ref{EB}), then he is less likely to guess $a_i$ correctly, and he will be less successful in biasing the string. This applies for each of the security conditions we defined. 

As we stated above, a further available strategy for a dishonest Bob is simply to play the protocol improperly, causing Alice to abort. He may do this at any time - for example he may do it near the end of the protocol if it seems that Alice's output string is not going to be to his liking. It is clear, however, that such a strategy cannot increase the average bias. If we recall that when using the Shannon entropy condition, we assume that Alice does not abort but instead outputs a random string, then it is also clear that this strategy cannot help Bob decrease the Shannon entropy of Alice's output. Thus we do not need to consider it in this case either (it was largely to avoid these complications that we adopted this convention). 

Eq.~(\ref{EB}), therefore, implies Eqs.~(\ref{noisyeb}) and (\ref{noisyhb}). With appropriate settings for $\theta$, Eq.~(\ref{EB}), along with our results for Alice below, implies Eqs.~(\ref{noiselesseaeb}) and (\ref{noiselesshahb}).

\subsection{Dishonest Alice: uncorrelated cheating}

In this section, we consider a single round of Protocol 1, and for
simplicity of notation we drop the subscript
$i$. Thus we denote by $a$ the bit sent by Alice at step 4 of the
protocol, by $b$ the bit sent by Bob at step 3, by $x$ and $y$ Alice's and Bob's outputs, and by $f$ the result of Bob's measurement. We 
denote by $E(f)$ the expectation value of $f$, i.e., the probability
that $f=1$. We prove the following lemma.
\begin{lemma}\label{singleroundlemma}
For a single round of Protocol 1, 
if Bob is honest then for any strategy of Alice we have the constraint
\begin{eqnarray}\label{ff}
\forall c\in \{0,1\}\quad  \mathrm{P}^{S_A H_B} (y=c) &\leq& \min \left\{
\frac12 + \frac{\sqrt{1-E(f)}}{\sqrt{2}\sin^2 \theta }+
\frac{1 - E(f)}{\sin^2 \theta} \ ,
\ 1\right\} \equiv {\cal F}(E(f)),          
\end{eqnarray}
where we define for future use 
${\cal F}(y) =  \min \left\{ \frac12 +
\frac{\sqrt{1-y}}{\sqrt{2} \sin^2 \theta } + \frac{1-y}{\sin^2\theta}\ ,\ 
1\right\}$, which is a concave
monotonically decreasing function for $x\in [0,1]$.
\end{lemma}
Clearly this lemma tells us that the more Alice cheats on a particular round, the more likely Bob's test on that particular round is likely to be failed.

{\bf Proof.}
Let us consider Alice's most
general strategy for a single round. This consists in Alice preparing a (possibly mixed) state
$\rho_{AB}$ and sending the $B$ subsystem to Bob via a noiseless
channel. Denote Bob's reduced density matrix by $\rho_B$.
Alice waits until she receives Bob's bit $b$. If $b=0$, she then performs a two
outcome POV measurement $M_0$ on subsystem A. Denote the two outcomes
$M_{00}$ and $M_{01}$. Alice declares $a=0$ if she
obtains outcome $M_{00}$ and declares $a=1$ if she
obtains outcome $M_{01}$. If $b=1$, she
performs a POV measurement $M_1$ with two outcomes $M_{10}$  and
$M_{11}$. Alice declares $a=1$ if she
obtains outcome $M_{11}$ and declares $a=0$ if she
obtains outcome $M_{10}$.
Suppose that Bob's reduced density matrices, conditioned on Alice getting outcomes
$M_{00}$, $M_{01}$, $M_{11}$, $M_{10}$, are $\sigma$, $\bar \sigma$,
$\tau$, $\bar \tau$ respectively, and denote by $q$ ($q'$) the
probability of obtaining outcome $M_{00}$ ($M_{11}$) if Alice
performs measurement $M_0$ ($M_1$). Then we can write
\begin{equation}
\rho_B = q \sigma + (1-q) \bar \sigma = q' \tau + (1-q') \bar \tau\ .
\label{rhoB}
\end{equation}
The expected (unnormalised) density matrix if Alice declares $a=0$ is
\begin{equation}
\rho_0 = q \sigma + (1 - q') \bar \tau 
\label{rho1}
\end{equation}
and the expected (unnormalised) density matrix if Alice declares $a=1$ is
\begin{equation}
\rho_1 = q' \tau + (1 - q) \bar \sigma \ .
\label{rho2}
\end{equation}
The expected fidelity $E(f)$ for this coin toss is 
\begin{equation}
E(f) = \frac12 (\langle \psi_0 | \rho_0 |\psi_0 \rangle 
+ \langle \psi_1 | \rho_1 |\psi_1 \rangle).
\label{fff}
\end{equation}

Let us now use the fact that there is an inherent symmetry in Protocol
1. Denote by $U_B$
the unitary transformation such that $U_B |\psi_0\rangle =
|\psi_1\rangle$ and $U_B |\psi_1\rangle = |\psi_0\rangle$. Suppose
that Alice prepares the state $I_A \otimes U_B \rho_{AB} I_A \otimes
U_B^\dagger$, sends Bob his part of this state, and carries out
measurement $M_1$ if $b=0$ (and Alice declares outcome
$a=0$ if she gets outcome $M_{11}$ and declares $a=1$ if she gets
outcome $M_{10}$), and measurement $M_0$ if $b=1$ (with
the role of the outcomes similarly permuted). In this strategy $q$ is
replaced by $q'$, $\sigma$ by $U_B \tau U_B^\dagger$, etc. This
strategy obviously gives Alice the same expected bias since 
Bob's bit $b$ is
random and initially unknown to Alice.

Alice could also randomly choose between these two strategies. This
will yield a symmetric strategy which will have the same expected bias as the
original strategy. We can describe this symmetric
strategy by the initial state
\[
\frac12
( |0\rangle \langle 0| \otimes \rho_{AB}
+ |1\rangle \langle 1| \otimes (1_A \otimes U_B ) \rho_{AB} (1_A
\otimes U_B^\dagger))
\]
where the additional qubit is the coin Alice tosses to decide which
strategy to use. Alice's measurement $M_0$ now consists of the two
elements
$|0\rangle\langle 0|\otimes M_{00} + |1\rangle \langle 1| \otimes
M_{11}$
and $|0\rangle\langle 0|\otimes M_{01} + |1\rangle \langle 1| \otimes
M_{10}$, and similarly for $M_1$.
For this symmetric strategy we can once more write Eqs.~(\ref{rhoB}), (\ref{rho1}), and (\ref{rho2}), but now we have the identities
\begin{eqnarray}
&q=q'\ ,&\label{qq'}\\
& \tau = U_B \sigma U_B^\dagger\ ,\ 
\bar \tau = U_B \bar \sigma U_B^\dagger,&
\end{eqnarray}
which imply that 
\begin{equation}  
\langle \psi_0 | \sigma | \psi_0 \rangle= \langle \psi_1 | \tau |
\psi_1 \rangle\ ,\ 
\langle \psi_0 |\bar  \sigma | \psi_0 \rangle= \langle \psi_1 |\bar \tau |
\psi_1 \rangle \ .
\label{sigma}
\end{equation}

In summary, Alice can use a symmetric strategy which does not
decrease her expected bias, but for which the relations (\ref{qq'})
and (\ref{sigma}) are obeyed. With this simplification we have
\begin{equation}
\label{EE0}
E(f)=\langle\psi_0|\rho_0|\psi_0\rangle = \langle\psi_1|\rho_1|\psi_1\rangle.
\end{equation}
The proof of Eq.~(\ref{ff}) now closely follows
the steps of the proof of Theorem~1 in
Ref.~\cite{BM}. 
First, from Eqs.~(\ref{rho1}), (\ref{qq'}) and (\ref{EE0}) we deduce that
\begin{equation}
q \langle \psi_0|\sigma|\psi_0\rangle
+ (1-q) \langle \psi_0|\overline \tau |\psi_0\rangle
= E(f),
\end{equation}
which in turn implies that
\begin{equation}
\langle \psi_0|\sigma|\psi_0\rangle
\geq 1 - \frac{1 - E(f)}{q},
\label{EE1}
\end{equation}
and that
\begin{equation}
\langle \psi_0|\overline \tau |\psi_0\rangle
\geq   1 - \frac{1 - E(f)}{1-q} \ .
\label{EE2}
\end{equation}
Now we introduce the quantity $D(\rho,\rho')=(1/2)\mathrm{Tr}\sqrt{(\rho-\rho')^{\dagger}(\rho-\rho')}$, where this is the trace distance between states $\rho$ and $\rho'$. We have that $D(\rho, \rho')\leq \sqrt{1 - F(\rho,\rho')}$
for arbitrary states $\rho$ and $\rho'$ (for this relation and others used below, see, e.g., Ref.~\cite{nielsenchuang}, although note the slightly different definition of fidelity). This gives
\begin{equation}
D(\sigma, |\psi_0\rangle) \leq \sqrt{\frac{1 - E(f)}{q}}
\quad , \quad
D(\overline \tau, |\psi_0\rangle) \leq \sqrt{\frac{1 - E(f)}{1-q}}\ .
\end{equation}
Using the same line of reasoning we can show that
\begin{equation}
D(\tau, |\psi_1\rangle) \leq \sqrt{\frac{1 - E(f)}{q}}
\quad , \quad
D(\overline \sigma, |\psi_1\rangle) \leq \sqrt{\frac{1 - E(f)}{1-q}}\ .
\label{EE14}
\end{equation}

Now we project Eq.~(\ref{rhoB}) onto $P=|\psi_0\rangle\langle\psi_0|$ to obtain
\begin{equation}
q {\rm Tr} (P \sigma)
+ (1-q){\rm Tr} (P \overline \sigma)
=
q {\rm Tr} (P \tau)
+ 
(1-q) {\rm Tr} (P \overline \tau) \ .
\label{EE4}
\end{equation}
We bound each term in Eq.~(\ref{EE4}) as follows:
\begin{enumerate}
\item From Eq.~(\ref{EE1}), the first term is bounded by ${\rm Tr} P \sigma \geq 1 -
(1-E(f))/q$.
\item To bound the second term we use Eq.~(\ref{EE14}) and the fact that
$D(\rho, \rho')= \max_P | {\rm Tr} (P \rho) - {\rm Tr}(P\rho')|$, where the maximum is over all projection operators $P$, in order to obtain
\begin{equation}
{\rm Tr} (P \overline \sigma)
\geq {\rm Tr} (P |\psi_1\rangle\langle \psi_1|)
- D(\psi_1,\overline \sigma)
\geq \cos^2 \theta
- \sqrt{\frac{1- E(f)}{1-q}}.
\end{equation}
\item
Similarly we get
\begin{equation}
{\rm Tr} (P \tau)
\leq {\rm Tr} (P |\psi_1\rangle\langle \psi_1|)
+ D(\psi_1,\tau )
\leq \cos^2 \theta
+ \sqrt{\frac{1- E(f)}{q}}.
\end{equation}
\item
Finally, ${\rm Tr}(P \overline \tau) \leq 1$.
\end{enumerate}
Together these inequalities imply that
\begin{equation}
(2q-1)\sin^2\theta
\leq \sqrt{1 - E(f)}\left (\sqrt{q} + \sqrt{1-q}\right)+ 1- E(f),
\end{equation}
which implies Eq.~(\ref{ff}). \quad$\Box$

\subsection{Dishonest Alice: bounds on average bias}

If Alice always uses the same strategy at each round of the protocol, 
then Eq.~(\ref{noisyea}) follows
directly from Eq.~(\ref{ff}). However, Alice need not follow
the same strategy during all coin tosses. She can modify her strategy
at round $i$ depending on what happened during the previous rounds. In
fact her most general strategy is to use quantum correlations: 
the quantum state she uses at
round $i$ can be entangled with the states she sent Bob during the
previous rounds. In this section we will show that such correlated
strategies can only help Alice marginally, and that the uncorrelated
cheating strategy described in the previous section is essentially optimal.

But can an entangled cheating strategy help Alice
at all? In principle, yes. Indeed Alice does not know the outcome of
Bob's measurement. By using states that are entangled over different rounds, Alice can obtain
some information about Bob's measurement and use this information to
modify her strategy during subsequent rounds. Analysing the effect of such
entangled strategies seems very difficult. So our approach will be to
modify the protocol in such a way that entanglement over rounds can no longer
help. Then we analyse the security of the modified protocol.
\\ \\
{\bf Protocol 2:}
The protocol is the same as Protocol 1, except that in step 5, Bob
carries out a complete measurement on the state sent by Alice, i.e., he
measures an orthonormal basis that includes
$|\psi_{a_i}\rangle$. He then reveals the result of the measurement to
Alice.
\\ \\
For this protocol the amount by which Bob can cheat is unchanged. But
Alice now knows everything that occurred at Bob's site. It is therefore
easier for Alice to cheat in Protocol~2 than in Protocol~1. On the
other hand, by carrying out a complete measurement and revealing the
result of the measurement, Bob has destroyed all entanglement that
could have existed between himself and Alice. Entanglement between
rounds therefore cannot help Alice in this protocol. She can, however, use the
information provided by Bob to correlate classically her strategy at
round $i$ with what happened during previous rounds.

In what follows we analyse the security of Protocol~2 with respect to
a cheating Alice. Let us recall that
\begin{eqnarray}
\frac12 +\epsilon_A &=& \max_{S_A ,{\bf c}}\ \frac{1}{n} \sum_i
\mathrm{P}^{S_AH_B}(y_i=c_i)\nonumber\\ 
&=&\max_{S_A ,{\bf c}}\ \frac{1}{n}\sum_i
\mathrm{P}^{S_AH_B}(a_i\oplus b_i=c_i\ \&\ \frac{1}{n}\sum_i f_i \geq f^*),
\label{eqBBB}
\end{eqnarray}
where we recall that $f_i=1$ if Bob 
finds outcome $|\psi_{a_i}\rangle$ when he carries out his measurement
at step 5 of the protocol, and $f_i=0$ otherwise.
In the second line we have 
rewritten the average bias as the probability that
Alice gets the outcome she wants at each round and that Bob does not
abort at the end of the protocol.  

This leads us to define variables $q_i({\bf c})\in\{0,1\}$ that are equal to
$1$ if $c_i = a_i \oplus b_i$ and equal to zero if $c_i \neq a_i
\oplus b_i$ (independently of whether or not Bob aborts at the end of
the protocol). We also define $I_{\mathrm{pass}}$ as equal to $1$ if Bob's test is passed and $0$ otherwise.
In terms of these variables we can rewrite Eq.~(\ref{eqBBB}) as
\begin{eqnarray}
\frac12 + \epsilon_A &=& \max_{S_A ,{\bf c}}\ \frac{1}{n}\sum_i \mathrm{E}
\left( q_i \ \times \ I_{\mathrm{pass}} \right)\nonumber\\ 
&=&\max_{S_A ,{\bf c}} \mathrm{E}\left( \left(\frac{1}{n}\sum_i q_i\right)\ \times\ I_{\mathrm{pass}} \right).
\label{eqBBB'}
\end{eqnarray}
In what follows we will obtain a lower bound on    
$(1/n)\sum_i \mathrm{P}(a_i\oplus b_i=c_i)$
that depends on $\frac{1}{n}\sum_i f_i$. From this bound we will
immediately deduce a bound on $\epsilon_A$.  

In general Alice's strategy may depend on
what happened during the previous rounds. Let us denote by $h_i$
all the events that occurred in round $i$. This includes the values of $a_{i},b_{i}$, the outcome of Bob's measurement in round $i$, and the outcomes of any probabilistic decisions made by Alice during round $i$. We abbreviate the sequence $h_1,\ldots,h_{i-1}$ by $past_i$. Thus we denote by
$\mathrm{E}(q_{i}|past_i)= \mathrm{P}(q_i=1|past_i)$  the probability that $q_i=1$ given
what happened during previous rounds. 
We denote $\mathrm{E}(f_{i}|past_{i})= \mathrm{P}(f_i=1 |past_{i})$ the probability that Bob will
find outcome $|\psi_{a_i}\rangle$ when he carries out his measurement
at step 5 of the protocol, given what happened during the previous
rounds.

The importance of the quantity $\mathrm{E}(f_{i}|past_{i})$ is that we can
use Lemma~\ref{singleroundlemma} to relate it to $\mathrm{E}(q_{i}|past_{i})$ by:
\begin{equation}
\mathrm{E}(q_i|past_i)\leq  {\cal F}(\mathrm{E}(f_i|past_i)),
\end{equation}
where ${\cal F}$ is defined in Lemma~\ref{singleroundlemma}.

In order to use this result to bound the average bias $\epsilon_A$, we
will use the theory of martingales \cite{Hoeffding}. 
\\ \\
{\bf Definition.} 
Consider random variables $S_1,\ldots,S_n$, and $X_1,\ldots,X_n$, such that $\mathrm{E}(|S_i|)<\infty$ for all $i$. The sequence $S_1,\ldots,S_n$ is a \emph{super-martingale} with respect to the sequence $X_1,\ldots,X_n$ if 
\begin{equation}\label{martingalecondition}
\mathrm{E}(S_i|X_1,\ldots ,X_{i-1}) \leq S_{i-1}.
\end{equation}
If the inequality is replaced with equality, then the sequence $S_1,\ldots,S_n$ is a $\emph{martingale}$ with respect to $X_1,\ldots,X_n$.
\\ \\
{\bf Hoeffding's inequality:}
Suppose that $S_1,\ldots,S_n$ is a super-martingale with respect to $X_1,\ldots,X_n$. Suppose also that $\mathrm{P}(|S_i-S_{i-1}|)\leq 1)=1$ for all $i$.
Hoeffding's inequality states that for all $l>0$,
\[
\mathrm{P}\left(\max_{1\leq K\leq n} S_K \geq l \sqrt{n}\right) \leq \exp(-l^2/2),
\]
i.e., the fluctuations of (super-)martingales cannot be much larger than those one
expects for independent random variables.
\\ \\
Now consider the variables $Y_i = f_i - \mathrm{E}(f_i|past_{i})$. The sequence $S_K=\sum_{i=1}^{K} Y_i$, for $K=1,\ldots,n$ is a martingale with respect to the sequence $h_1,\ldots,h_n$. The conditions in
Hoeffding's inequality are obeyed. Hence we have that for any strategy of Alice,
\begin{equation}
\mathrm{P}\left(\frac{1}{n}\sum_{i=1}^n f_i - \mathrm{E}(f_i|past_{i}) \geq \frac{l}{\sqrt{n}}\right) \leq \exp(-l^2/2).
\label{boundf}\end{equation}
This expresses the fact that the actual results of Bob's measurements, given by
$\sum_i f_i$, cannot differ much from their expected value given
the past.

The variables $S'_K=\sum_{i=1}^K Z_i$, where $Z_i= q_i - {\cal F}(\mathrm{E}(f_i|past_{i}))$ are a super-martingale with respect to $h_1,\ldots,h_n$ and also obey the conditions in Hoeffding's inequality, with
$\mathrm{E}(Z_i|past_{i})\leq 0$ and $-1 \leq Z_i \leq +1$. Hence
\begin{equation}
\mathrm{P}\left(\frac{1}{n}\sum_{i=1}^n q_i - {\cal F}(\mathrm{E}(f_i|past_{i})) \geq \frac{l}{\sqrt{n}}\right) 
\leq \exp(-l^2/2).
\label{boundq}\end{equation}
This expresses the fact that the average of Alice's results cannot
exceed by much the average of the bounds given by Lemma~\ref{singleroundlemma}.

Concavity of ${\cal F}$ implies that
\[
{1\over n}\sum_{i=1}^n {\cal F}(\mathrm{E}(f_i|past_{i})) \leq 
{\cal F}\left({1\over n}\sum_{i=1}^n \mathrm{E}(f_i|past_{i})\right).
\]
Inserting this in Eq.~(\ref{boundq}) yields
\begin{equation}
\mathrm{P}\left({1\over n}\sum_{i=1}^n q_i \geq  
{\cal F}\left({1\over n}\sum_{i=1}^n \mathrm{E}(f_i|past_{i}) 
\right) + {l \over \sqrt{n}}\right) \leq \exp(-l^2/2).
\label{boundq2}\end{equation}

Using the union bound for Eqs.~(\ref{boundf}) and (\ref{boundq2}) and the
fact that ${\cal F}$ is a decreasing function, one has
\begin{eqnarray*}
&&\mathrm{P}\left({1\over n}\sum_{i=1}^n q_i \geq  
{\cal F}\left({1\over n}\sum_{i=1}^n f_i -
{l \over \sqrt{n}} \right) + {l \over \sqrt{n}}\right)
\leq 2 \exp(-l^2/2)\label{boundq3}\\
&\Rightarrow &\mathrm{P}\left({1\over n}\sum_{i=1}^n q_i \geq  
{\cal F}\left({1\over n}\sum_{i=1}^n f_i -
{l\over\sqrt{n}} \right) + {l\over\sqrt{n}} \ \& \ \mathrm{pass}\right)
\leq 2 \exp(-l^2/2)\\
&\Rightarrow&\mathrm{P}\left({1\over n}\sum_{i=1}^n q_i \geq  
{\cal F}\left(f^*-{l\over\sqrt{n}} \right) + {l\over\sqrt{n}} \ \&\ \mathrm{pass}\right)
\leq 2 \exp(-l^2/2).
\end{eqnarray*}

We now denote the event that
\[
\frac{1}{n} \sum_{i=1}^n q_i \geq {\cal F}\left(f^*-\frac{l}{\sqrt{n}} \right) + \frac{l}{\sqrt{n}}
\]
as $J$. We define $I_J$ such that $I_J=1$ if $J$ occurs and $0$ otherwise, and $I_{\bar{J}}=1-I_J$.
We go back to Eq.~(\ref{eqBBB'}), which we write as
\begin{eqnarray}
\frac12 +\epsilon_A &=& \max_{S_A,{\bf c}} \left[ \mathrm{E}\left( \left(\frac{1}{n}\sum_i q_i\right)\ \times\ I_{\mathrm{pass}}\ \times\ I_J \right) + \mathrm{E}\left( \left(\frac{1}{n}\sum_i q_i\right)\ \times\ I_{\mathrm{pass}} \ \times\ I_{\bar{J}} \right)\right]\nonumber\\
&\leq & 2 \exp (-l^2/2) + {\cal F}\left(f^*-\frac{l}{\sqrt{n}} \right) + \frac{l}{\sqrt{n}}.\label{boundq4B}
\end{eqnarray}


Taking $f^*=1$, $\sin^2\theta=(\ln n)^{1/6}n^{-1/6}$, and
$l=\sqrt{\ln n}$ yields 
\begin{equation}
\epsilon_A \leq \frac{1}{\sqrt{2}}\left(\frac{\ln n}{n}\right)^{1/12} + \left(\frac{\ln n}{n}\right)^{1/3} + \sqrt{\frac{\ln n}{n}}+\frac{2}{\sqrt{n}},
\end{equation}
which, along with our results for Bob above, gives Eq.~(\ref{noiselesseaeb}). On the other hand, if we take $f^*<1$, with $\theta$ fixed, and $l=\sqrt{\ln n}$, we get
\begin{equation}
\epsilon_A \leq \frac{\sqrt{(1-f^*)}}{\sqrt{2}\sin^2\theta} + \frac{1-f^*}{\sin^2\theta} + \sqrt{\frac{\ln n}{n}}\left(\frac{1}{2\sqrt{2}\sin^2\theta\sqrt{1-f^*}} + \frac{1}{\sin^2\theta} + 1\right) + \frac{2}{\sqrt{n}},
\end{equation}
which gives Eq.~(\ref{noisyea}). 

\subsection{Dishonest Alice: bounds on entropy}

Finally, we prove a lower bound on the entropy of Bob's output, which will give us
Eqs.~(\ref{noisyha}) and (\ref{noiselesshahb}). In this section, we define the string ${\bf c}$ such that $c_i=a_i\oplus b_i$.
We begin by defining the set $S$, which is a subset of all possible strings ${\bf c}$:
\begin{equation}\label{sdefinition}
S = \left\{{\bf c} : \frac{1}{n}\sum_{i=1}^{n}\mathrm{E}(f_i|c_1,\ldots,c_{i-1})\geq f^* - \frac{l}{\sqrt{n}} \right\}.
\end{equation}
The idea is that conditioned on passing Bob's test, the probability that ${\bf c}\in S$ is high, and that this can be used to bound the entropy of ${\bf y}$ conditioned on passing the test. On the other hand, if the test is failed, Bob will simply output a random string, so that the entropy conditioned on this event is also high (equal to $n$, in fact). 

Eq.~(\ref{boundf}) reads
\[
\mathrm{P}\left(\frac{1}{n}\sum_{i=1}^n f_i - \mathrm{E}(f_i|past_{i}) > \frac{l}{\sqrt{n}}\right) \leq \exp(-l^2/2), 
\]
which implies
\[
\mathrm{P}({\bf c}\in \bar{S} \ \&\ \mathrm{pass}) \leq \exp(-l^2/2),
\]
where $\bar{S}$ denotes the complement of $S$. Thus
\begin{equation}\label{sbound}
\mathrm{P}({\bf c}\in S|\mathrm{pass}) \geq 1 - \frac{\exp(-l^2/2)}{\mathrm{P}(\mathrm{pass})}.
\end{equation}
We shall use this below.

We bound the probability of a particular string ${\bf c}$, where ${\bf c}\in S$. Write
\begin{eqnarray}
\mathrm{P}({\bf c} \ \&\ \mathrm{pass})\leq \mathrm{P}({\bf c})
&=&\prod_{i=1}^{n} \mathrm{P}(c_i|c_1,\ldots,c_{i-1})\nonumber\\
&\leq & \prod_{i=1}^{n} {\cal F}(\mathrm{E}(f_i|c_1,\ldots,c_{i-1}))\nonumber\\
&\leq & \left[\frac{1}{n}\sum_{i=1}^{n} {\cal F}(\mathrm{E}(f_i|c_1,\ldots,c_{i-1}))\right]^n\nonumber\\
&\leq &\left[{\cal F}\left(\frac{1}{n} \sum_{i=1}^{n} \mathrm{E}(f_i|c_1,\ldots,c_{i-1})\right)
\right]^n,\nonumber\\
\end{eqnarray}
where we have used the fact that $\prod_{i=1}^{n} x_i \leq (\sum_{i=1}^{n} x_i/n)^n$, and the concavity of ${\cal F}$. 
Now, if ${\bf c}\in S$, then using Eq.~(\ref{sdefinition}) we immediately deduce that
\begin{equation}
\mathrm{P}({\bf c}) \leq [{\cal F}(f^* - l/\sqrt{n})]^n.
\end{equation}
This implies
\begin{equation}
\mathrm{P}({\bf c}|\mathrm{pass})\leq \frac{[{\cal F}(f^*-l\sqrt{n})]^n}{\mathrm{P}(\mathrm{pass})}.
\end{equation}
Using the fact that if Bob's test is passed then ${\bf y}={\bf c}$, we have
\begin{eqnarray}
H(\mathrm{P}({\bf y}|\mathrm{pass}))=H(\mathrm{P}({\bf c}|\mathrm{pass}))&=& - \sum_{\bf c} \mathrm{P}({\bf c}|\mathrm{pass})\log \mathrm{P}({\bf c}|\mathrm{pass})\nonumber\\
&\geq & - \sum_{{\bf c}\in S}\mathrm{P}({\bf c}|\mathrm{pass})\log \mathrm{P}({\bf c}|\mathrm{pass})\nonumber\\
&\geq & - \mathrm{P}({\bf c}\in S|\mathrm{pass}) \log \frac{[{\cal F}(f^* - l/\sqrt{n})]^n}{\mathrm{P}(\mathrm{pass})}.
\end{eqnarray}
Using Eq.~(\ref{sbound}), we get
\begin{equation}\label{result}
H(\mathrm{P}({\bf y}|\mathrm{pass}))\geq \left( 1 - \frac{\exp(-l^2/2)}{\mathrm{P}(\mathrm{pass})}\right)\left(-n\log[{\cal F}(f^* - l/\sqrt{n})] + \log\mathrm{P}(\mathrm{pass})\right).
\end{equation}
Finally,
\begin{eqnarray}
H(\mathrm{P}({\bf y})) &\geq& \mathrm{P}(\mathrm{pass}) H(\mathrm{P}({\bf y})|\mathrm{pass}) + (1-\mathrm{P}(\mathrm{pass})) H(\mathrm{P}({\bf y})|\mathrm{abort})\nonumber\\
&\geq& \mathrm{P}(\mathrm{pass})\left( 1 - \frac{\exp(-l^2/2)}{\mathrm{P}(\mathrm{pass})}\right)\Big(-n\log[{\cal F}(f^* - l/\sqrt{n})] + \log\mathrm{P}(\mathrm{pass})\Big)+ (1-\mathrm{P}(\mathrm{pass})) n\label{hyboundwithppass}.
\end{eqnarray}
Now, considering $\mathrm{P}(\mathrm{pass})$ as an independent variable in Eq.~(\ref{hyboundwithppass}), with $0\leq \mathrm{P}(\mathrm{pass}) \leq 1$, it is easy to show that the right hand side is minimised if $\mathrm{P}(\mathrm{pass})=1$. (Note that we are only dealing with a lower bound, so this does not imply that Alice's best strategy will have $\mathrm{P}(\mathrm{pass})=1$.) We get
\begin{equation}
H(\mathrm{P}({\bf y})) \geq -n\log[{\cal F}(f^* - l/\sqrt{n})]\left( 1 - \exp(-l^2/2) \right).
\end{equation}
Setting $f^*=1$, $l=\sqrt{\ln n}$ and $\sin^2\theta=n^{-1/8}(\ln n)^{1/8}$ gives
\begin{equation}
H(\mathrm{P}({\bf y})) \geq -n\log\left[\frac12+\frac{(\ln n)^{1/4}}{\sqrt{2}(\sin^2\theta) n^{1/4}} + \frac{\sqrt{\ln n}}{\sqrt{n}\sin^2\theta}\right]\left(1-\frac{1}{\sqrt{n}}\right),
\end{equation}
which, along with our results for Bob above, gives Eq.~(\ref{noiselesshahb}). On the other hand, if we set $f^*<1$, with constant $\theta$ and $l=\sqrt{\ln n}$, then we get
\begin{equation}
H(\mathrm{P}({\bf y})) \geq -n\log\left[\frac12+\frac{\sqrt{1-f^*+l/\sqrt{n}}}{\sqrt{2}\sin^2\theta} + \frac{1-f^*+l/\sqrt{n}}{\sin^2\theta}\right]\left(1-\frac{1}{\sqrt{n}}\right),
\end{equation}
which gives Eq.~(\ref{noisyha}).

\section{Further discussion}\label{discussionsection}

In this section, we discuss a few further points of interest.

\subsection{Bit-string generation and bit-string commitment}

It is well known that there are relationships between the cryptographic tasks of coin tossing and bit commitment. Briefly, the idea in bit commitment is that Alice must commit a bit to Bob in such a manner that Bob cannot determine its value. At a later stage, Alice reveals the bit to Bob, and must not be able to reveal a value different from that which she committed. It is clear that a secure bit commitment implies secure (strong) coin tossing: Alice commits a bit to Bob, who guesses it's value, and Alice then reveals whether Bob was correct. The outcome of the coin toss is $0$ if Bob was correct and $1$ otherwise. Bit commitment can obviously be generalised to bit-string commitment, and secure bit-string commitment will imply secure bit-string generation in a similar manner. 

It was originally discovered by Lo and Chau \cite{lochau}, and independently by Mayers \cite{mayers}, that quantum bit commitment is not possible with arbitrary security. Kitaev's bound for quantum coin tossing provides an alternative proof of this fact. Spekkens and Rudolph have investigated partially secure quantum bit commitment \cite{spekkensrudolph} and Kent has introduced a scheme for quantum bit-string commitment with partial security \cite{kentbitstringcommitment}.

We note that our protocol cannot be regarded as a bit-string commitment scheme, due to its sequential nature. It may seem tempting to modify the protocol so that i) Alice sends
$|{\psi_{a_1}}\rangle,\ldots,|{\psi_{a_n}}\rangle$ to Bob, ii) Bob sends
$b_1,\ldots,b_n$, and iii) Alice sends $a_1,\ldots,a_n$. This modified protocol is essentially a bit-string commitment protocol with a bit-string generation protocol built on top (the former being similar to Kent's protocols for quantum bit-string commitment).

Unfortunately, however, in the modified protocol there are new cheating strategies for Alice. Alice may prepare an entangled state of $n+1$ systems, keeping $1$ and sending $n$ to Bob. She then waits for $b_1,\ldots,b_n$ before performing a measurement on her system. The values of $a_1,\ldots,a_n$ may depend on the outcome of this measurement, and on the values of all of $b_1,\ldots,b_n$. It can be shown that if Alice uses such a strategy, then even in the noiseless case, the protocol is not relatively secure (in the sense defined in Sec.~\ref{securitysection}). In fact, the security is not significantly better than the trivial classical protocol of Sec.~\ref{classicalsection}. Our notion of relative security can be adapted to bit-string commitment, and it follows that the associated bit-string commitment is not relatively secure either. These remarks apply equally to Kent's protocols for bit-string commitment (this does not contradict any of Kent's results, as his proofs involve weaker notions of security). We conjecture that no quantum bit-string commitment protocol with relative security exists.   

\subsection{Improving the protocol}

In Protocol~1 as described, Bob is fairly restricted in that he only estimates the fidelity of the states sent by Alice, and aborts if this is too low. In realistic situations, however, Bob may have a good idea of what shape the noise should have in the absence of cheating. For example, he may know that in the absence of cheating, the channel employed is a depolarising channel. In this case, Bob can perform quantum tomography on the states sent by Alice (separately for those rounds with $a_i=0$ and $a_i=1$). Bob will abort if the states received are not close to the noisy versions of the states sent he is expecting. This will obviously restict possible cheating by Alice, as she must reproduce the actual noisy states expected by Bob and not merely something with equivalent fidelity. In Ref.~\cite{BM}, the case was investigated in which Bob performs tomography, the channel is a depolarising channel, and Alice is restricted to individual attacks. The result, expressed in Theorem~2 of Ref.~\cite{BM}, indicates that security against Alice in this case may be significantly improved. It may be, however, that the required tomographic measurements are practically difficult to perform in a given implementation.  

\subsection{Classical post-processing}
In the case of quantum key distribution in the presence of noise, a potential eavesdropper may have partial knowledge of the raw key. In the absence of quantum error correction techniques, the honest parties may use classical privacy amplification of the raw key data in order to reduce the eavesdropper's knowledge. If the noise is not too great, then an arbitrarily secure key may be obtained in this way \cite{shorpreskill}. In light of this, it is natural to ask whether some kind of classical post-processing could improve the security of the bit-string generated by our protocol, at the expense of reducing the length of the string. From Kitaev's bound, we know that arbitrary security will not be possible, but could a relatively secure string be generated from a partially secure string? Or the level of partial security improved? 

In fact, although we do not offer a proof, we are sceptical.  Classical information processing in general offers rather limited possibilities for two mistrustful parties, as opposed to two honest parties trying to defeat an eavesdropper. It is easy to see that certain ideas, such as taking parities of subsets of the string, do not work. The essential problem is that if a classical post-processing scheme requires randomness, then a cheating party will be able to bias it. Generating randomness trusted by both parties is the problem of bit-string generation in the first place. 

\begin{acknowledgments}
We would like to thank Richard Gill for introducing us to Hoeffding's inequality, and Renato Renner for useful discussions on classical bit-string generation. We acknowledge financial support from the Communaut\'{e} Fran\c{c}aise de Belgique under Grant No. ARC 00/05-251, from the IUAP programme of the Belgian Government under Grant No. V-18, and from the EU under project RESQ (IST-2001-37559).
\end{acknowledgments}

\end{document}